\documentclass[twoside,letterpaper]{mn2eR}  
  
\newcommand{\ltsima} {$\; \buildrel < \over \sim \;$} 
\newcommand{\simlt}  {\lower.5ex\hbox{\ltsima}}            
\newcommand{\gtsima} {$\; \buildrel > \over \sim \;$} 
\newcommand{\simgt}  {\lower.5ex\hbox{\gtsima}}            

\newcommand{\rsun}{$R_\odot$}
\newcommand{\lsun}{$L_\odot$}

\title[Optical flickering of RS Oph]
{UBVRI observations of the flickering of RS Ophiuchi at Quiescence   
\thanks{based on data from Bulgarian observatories Rozhen and Belogradchik} }  
\author[Zamanov, Boeva, Bachev, et al. ]  
{R. K. Zamanov$^{1}$\thanks{e-mail: rkz@astro.bas.bg; sboeva@astro.bas.bg; bachevr@astro.bas.bg}, 
S. Boeva$^{1}$, R. Bachev$^{1}$, M. F. Bode$^{2}$,  D. Dimitrov$^{1}$,
\newauthor  K. A. Stoyanov$^{1}$, A. Gomboc$^{3}$, S. V. Tsvetkova$^{1}$, L. Slavcheva-Mihova$^{1}$,
\newauthor B. Spassov$^{1}$, K. Koleva$^{1}$, B. Mihov$^{1}$  \\\\  
$^{1}$ Institute of Astronomy, Bulgarian Academy of Sciences,   
       72 Tsarighradsko Shousse Blvd., 1784 Sofia, Bulgaria \\
$^{2}$ Astrophysics Research Institute, Liverpool John Moores University, 
Twelve Quays House, Birkenhead, CH41 1LD, UK \\  
$^{3}$  Faculty of Mathematics and Physics, University of Ljubljana, 
     Jadranska 19, 1000 Ljubljana, Slovenia  \\ 
}  
  
\begin{document}  
  
\date{Accepted  .. .. .. ... Received  ..  ..  2009}  
  
\pagerange{\pageref{firstpage}--\pageref{lastpage}} \pubyear{2009}  
  
\maketitle  
  
\label{firstpage}  
  
\begin{abstract}
 We report observations of the flickering variability of 
 the recurrent nova RS Oph at quiescence on the basis of
 simultaneous observations in  5 bands $(UBVRI)$.  
 RS~Oph has flickering source with 
 $(U-B)_0=-0.62 \pm 0.07$, $(B-V)_0=0.15 \pm 0.10$, $(V-R)_0=0.25 \pm 0.05$.

 We find for the flickering source  a temperature $T_{fl} \approx 9500 \pm 500 $~K,
 and luminosity $L_{fl}\sim 50 - 150$~$L_\odot$ (using a distance of $d=1.6$~kpc).
 
 We also find that on a $(U-B)$ vs $(B-V)$ diagram the flickering of 
 the symbiotic stars differs from that of 
 the cataclysmic variables.
 The possible source of the flickering is discussed. 
 
 The data are available upon request from the authors and on the web
 www.astro.bas.bg/$\sim$ rz/RSOph.UBVRI.2010.MNRAS.tar.gz. 
\end{abstract}  
  
  
\section{Introduction}                                                                                             
In the symbiotic recurrent nova RS Ophiuchi (HD 162214),
a near-Chandrasekhar-mass white dwarf (WD) accretes material from
a red giant companion (e.g., Hachisu \& Kato 2001; Sokoloski
et al. 2006). It experiences nova eruptions approximately
every 20 yr (Evans et al. 2008), with the most recent eruption having
occurred on 2006 February 12 (Narumi et al. 2006). 
Fekel et al. (2000) found that RS Oph has an orbital period of 455 days 
and give red giant and white dwarf (WD) masses of $2.3 \: M_{\odot}$ and close to $1.4 \: M_{\odot}$ 
respectively with a separation between the components of $a = 2.68 \times 10^{13}$ cm. 
For the range of spectral types suggested (Worters et al. 2007) for the red giant 
in the RS Oph system, its radius is smaller than its Roche lobe, 
and accretion onto the WD may occur only from the red giant wind.

The flickering (stochastic light variations on timescales of
a few minutes with amplitude of a few$\times0.1$ magnitudes)
is a variability observed in the three main types of binaries that contain white dwarfs 
accreting  material from a companion mass-donor star:  
cataclysmic variables (CVs), supersoft X-ray binaries, 
and symbiotic stars (Sokoloski 2003). 
The flickering of RS~Oph has been detected by Walker (1977). 
The systematic searches  for  flickering variability 
in  symbiotic stars and related objects (Dobrzycka et al. 1996a; 
Sokoloski, Bildsten \& Ho \ 2001;  Gromadzki et al. 2006)
have shown that among $\sim200$ symbiotic stars known, only
8 present flickering -- RS Oph, T CrB, MWC 560, Z And, 
V2116 Oph, CH Cyg, RT Cru, o Cet and V407 Cyg.

Here we investigate the flickering variability of RS~Oph in the $UBVRI$ bands
and discuss its possible origin.  

\begin{table*} 
\caption{CCD observations of RS Oph. In the table are given as follows:
the telescope, band, UT-start and UT-end of the run, exposure time, number of 
CCD images obtained, average magnitude in the corresponding band, 
minimum -- maximum magnitudes in each band, standard
deviation of the mean, typical observational error.}
\begin{center}
\begin{tabular}{llrrrrcrrccccrlcr}
\hline

telescope& band       &   UT        & Exp-time  &  N$_{pts}$  &  average & min-max     & stdev  & err   \\
      &               & start-end   & [sec]     &             & [mag]    & [mag]-[mag] & [mag]  & [mag] \\
\hline
{\bf 2008 July 6}  &   & JD2454654    \\
   2.0 m Rozhen	   & $U$ &19:27 - 21:26& 300       &  20 & 12.546 & 12.293 -- 12.721 & 0.103 &  0.011 \\
  50/70 cm Schmidt & $B$ &19:46 - 21:29& 100,60,20 &  50 & 12.471 & 12.380 -- 12.615 & 0.051 &  0.005 \\
   2.0 m Rozhen	   & $V$ &18:54 - 21:29&   30      & 196 & 11.215 & 11.079 -- 11.333 & 0.052 &  0.007 \\
   60 cm Rozhen    & $R$ &19:36 - 21:33&   15,20   & 301 & 10.277 & 10.172 -- 10.362 & 0.036 &  0.002 \\ 
   60 cm Belogr    & $I$ &19:42 - 21:22&   60      &  74 &  8.990 &  8.855 --  9.098 & 0.057 &  0.100 \\
\hline 
{\bf 2009 July 21} &  & JD2455034 \\

 50/70 cm Schmidt  & $U$ &20:58 - 21:52&   120     &  21  &  12.020  & 11.863 -- 12.166 &  0.092   &  0.013 \\
 60 cm Rozhen  	   & $B$ &20:17 - 21:38&   40      &  70  &  12.030  & 11.876 -- 12.277 &  0.098   &  0.007 \\
 60 cm Belogr      & $V$ &20:12 - 21:31&   30      &  81  &  11.009  & 10.891 -- 11.236 &  0.092   &  0.004 \\
 60 cm Belogr 	   & $R$ &20:12 - 21:31&   10      &  80  &  10.219  & 10.121 -- 10.406 &  0.074   &  0.003 \\
 60 cm Rozhen      & $I$ &20:17 - 21:38&   10      &  70  &   9.172  &  9.070 -- 9.325  &  0.066   &  0.003 \\
\hline

{\bf  2009 July 23}&  & JD2455036  \\ 
   2.0 m Rozhen    & $U$ &19:23 - 21:14&    120    &  41  &  11.659  & 11.482 -- 11.845 &  0.081   & 0.028   \\
 50/70 cm Schmidt  & $B$ &20:53 - 21:21&     30    &  46  &  11.978  & 11.857 -- 12.099 &  0.062   & 0.007   \\
   2.0 m Rozhen    & $V$ &19:24 - 21:15&     10    & 467  &  10.932  & 10.766 -- 11.093 &  0.070   & 0.008   \\
   60 cm Rozhen    & $I$ &19:40 - 21:17&      5    & 766  &   9.096  &  8.970 -- 9.216  &  0.050   & 0.005   \\
   60 cm Belogr    & $I$ &19:55 - 21:24&     10    & 404  &   9.118  &  8.998 -- 9.206  &  0.048   & 0.003   \\

\hline

\end{tabular}
\end{center}
\label{Tab1}
\end{table*}

\section{Observations}

 \begin{figure}
 \vspace{12.0cm}  
  \includegraphics{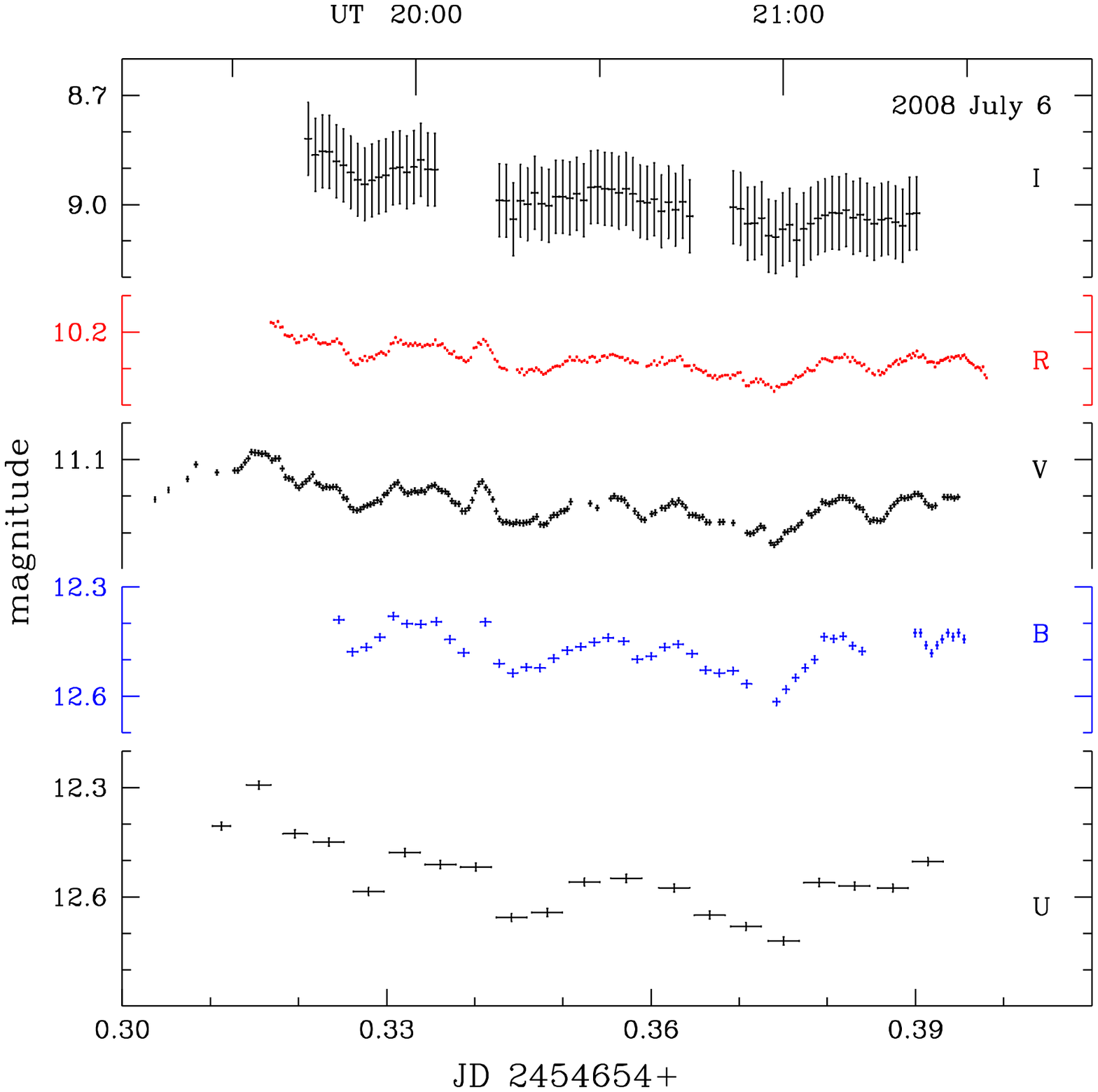}   
  \caption[]{Variability of RS Oph in the $UBVRI$ bands on 2008 July 6.
  }		    
\label{fig1-1}     
\end{figure}	    
 \begin{figure}
 \vspace{12.0cm}  
  \includegraphics{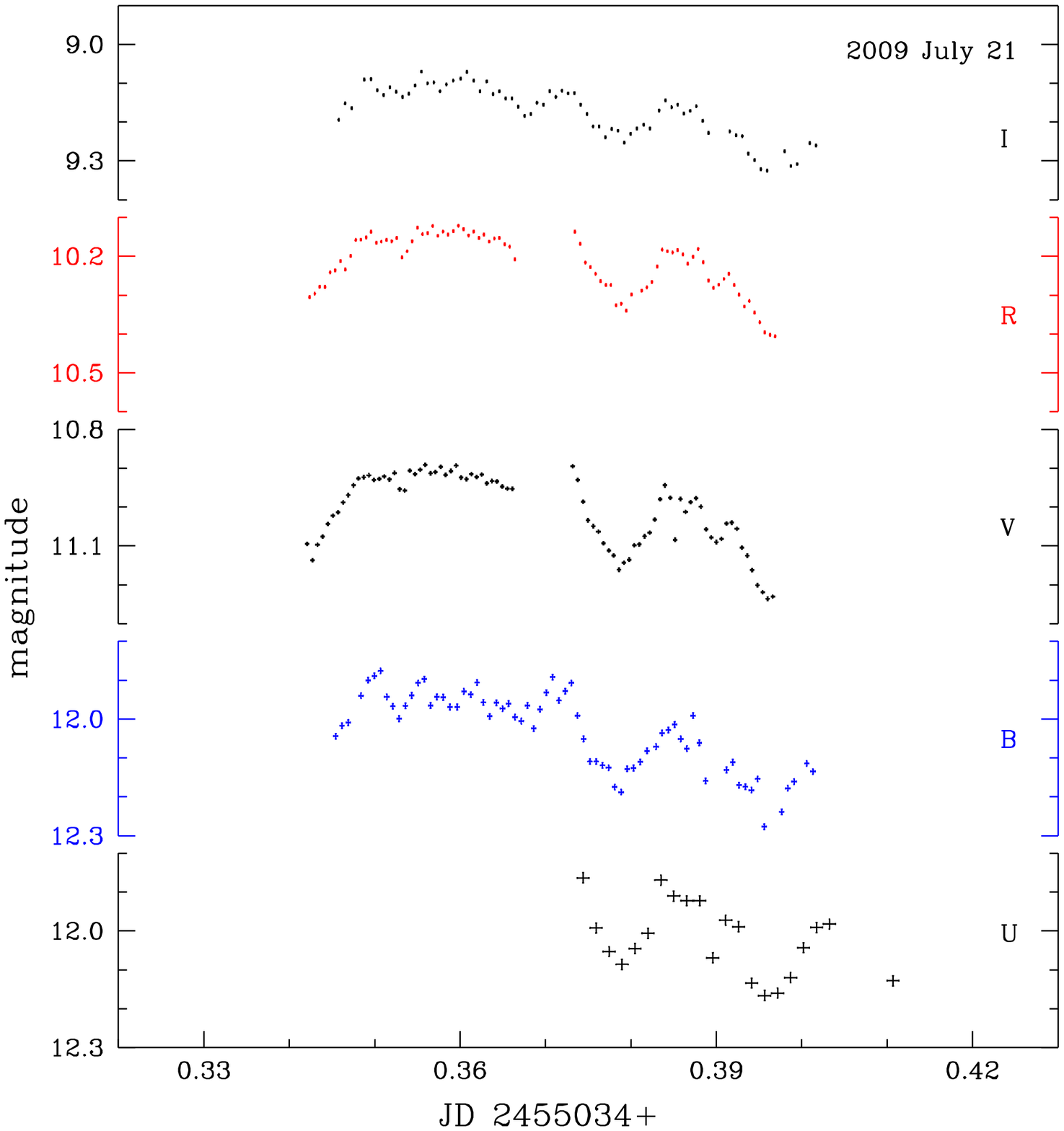}   
  \caption[]{Variability of RS Oph in the $UBVRI$ bands on 2009 July 21.
  }		    
\label{fig1-2}     
\end{figure}	    

 \begin{figure}
 \vspace{12.0cm}  
  \includegraphics{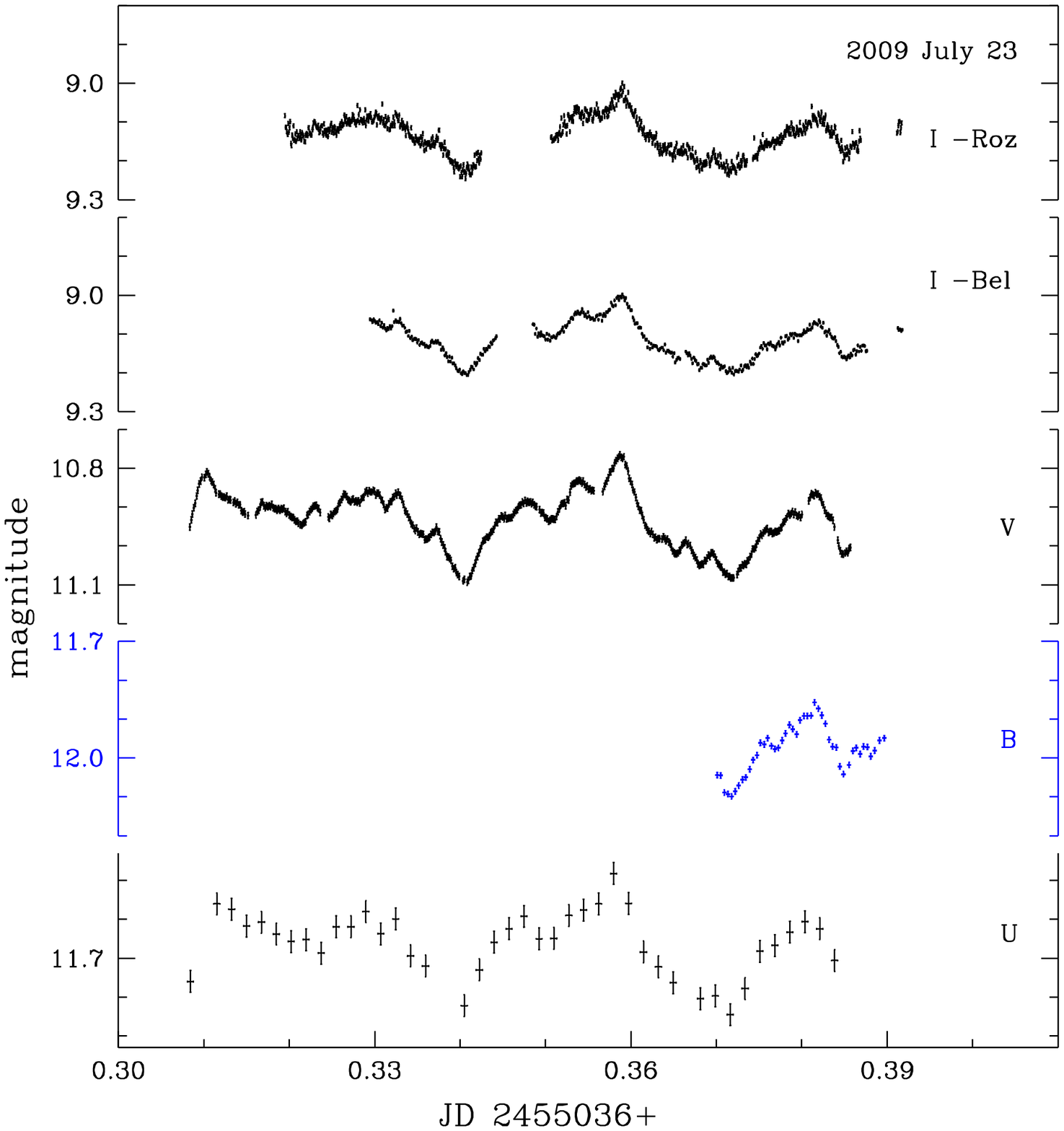}   
  \caption[]{Variability of RS Oph in the $UBVI$ bands on 2009 July 23.
  }		    
\label{fig1-3}     
\end{figure}	    

On the night of 2008 July 6, we observed RS Oph 
simultaneously with four telescopes equipped with CCD cameras.
The 2m RCC telescope of the National Astronomical Observatory  Rozhen 
equipped with a dual channel focal reducer observed 
in $U$ and $V$ bands.  In the $U$ band a
Photometrics CCD (1024x1024 px, field of view 4.9'x4.9') 
has been used, and 
in the $V$ band a VersArray 1330B (1340x1300, 6.3'x6.3'). 
The 60 cm Rozhen telescope observed in the $R$ band
(equipped with a FLI PL09000 CCD with 3056 x 3056 pixels and 5.7'x5.7');
the 50/70 cm Schmidt telescope of NAO Rozhen in the $B$ band 
(SBIG STL11000M CCD, 4008x2672 px, 16'x24'), 
and the 60 cm telescope of the Belogradchick Astronomical Observatory 
in the $I$ band (SBIG ST8 CCD, 1530x1020px, 6.4'x4.2'). 

On the night of 2009 July 21,  we observed RS Oph 
simultaneously with 3 telescopes.  
The 50/70 cm Schmidt telescope observed  in the $U$ band,
the 60 cm Rozhen telescope -- repeating $B$ and $I$ bands, and 
the 60 cm Belogradchick telescope --  repeating $V$ and $R$ bands.

On the night of 2009 July 23,  we observed RS Oph 
simultaneously with 3 telescopes.  
The 2m RCC telescope observed simultaneously in $U$ and $V$ bands, the 
50/70 cm Schmidt telescope -- in $B$, and the 60 cm Rozhen and Belogradchick telescopes -- in $I$ band.

All the CCD images have been bias subtracted, flat fielded, and standard 
aperture photometry has been performed. The data reduction and aperture photometry 
are done with IRAF and have been checked with alternative software packages. 
The comparison stars of Henden and Munari (2006) have been used.  

The results of our observations are summarized in Table~\ref{Tab1} and 
plotted in Figs.\ref{fig1-1}, \ref{fig1-2} and \ref{fig1-3}.
For each run  we measure the minimum, maximum, and average brightness
in the corresponding band, plus the standard deviation of the run.

\section{Colour variations}
As can be seen in Figs. \ref{fig1-1}, \ref{fig1-2}, \ref{fig1-3} and Table~1, during our observations
RS~Oph exhibited variability on a time scale of 1-30 minutes 
with amplitude 0.3 mag in $V$. The amplitude increases
to shorter wavelengths  and decreases to longer.

 \begin{figure} 
 \vspace{11.0cm}  
  \includegraphics{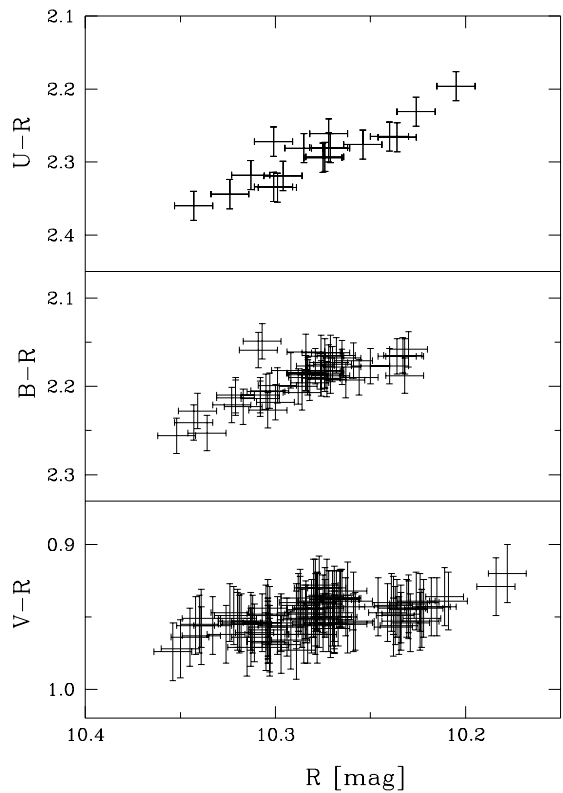}   
  \caption[]{$(U-R)$, $(B-R)$, and $(V-R)$  colours of RS Oph versus $R$ band magnitude for 
  2008 July 6.
  }		    
\label{figCOLORS}     
\end{figure}	    

\subsection{Magnitude - Colour relation}
 \begin{figure*}
 \mbox{}  
 \vspace{14.7cm}  
  \includegraphics{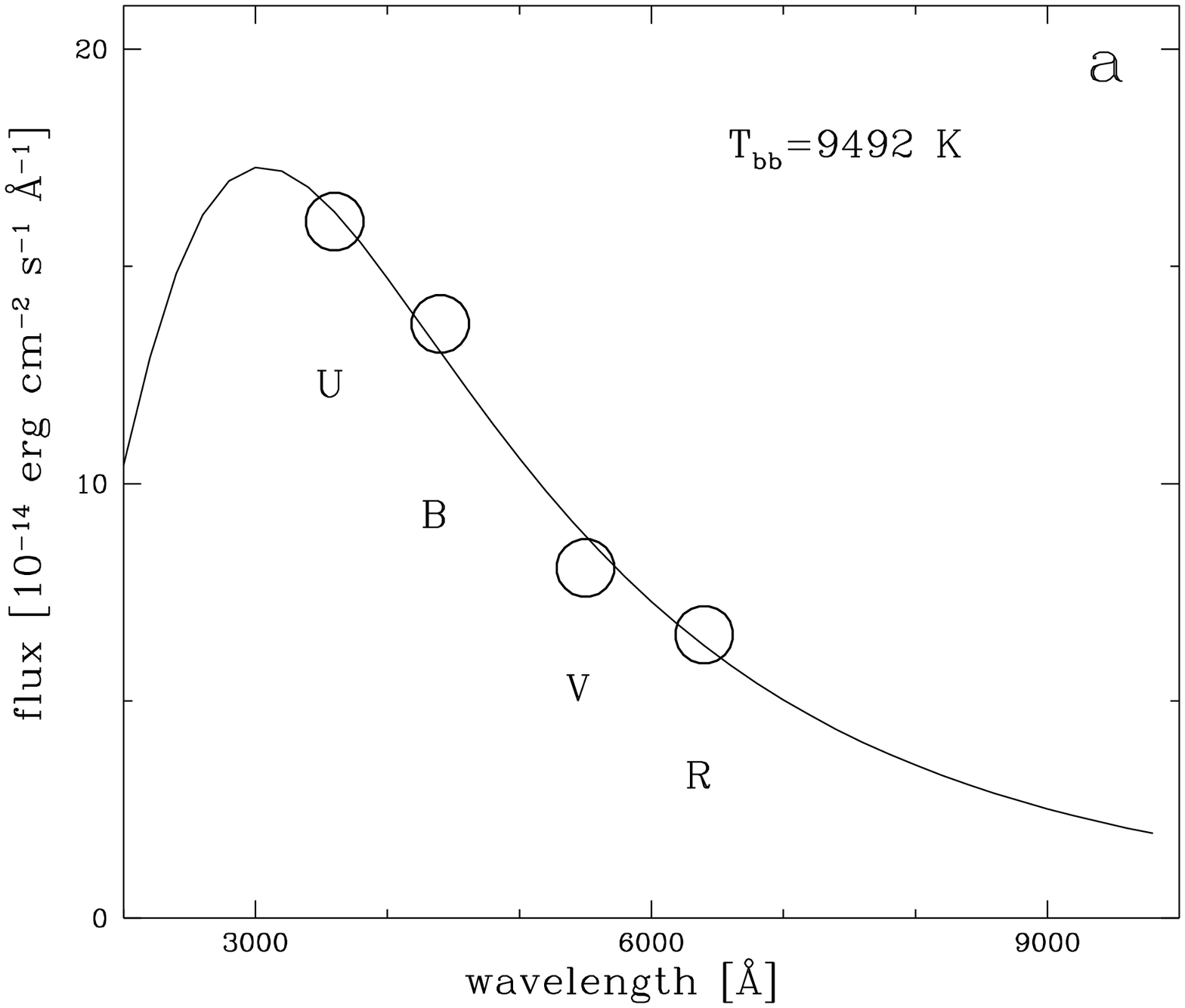}   
  \includegraphics{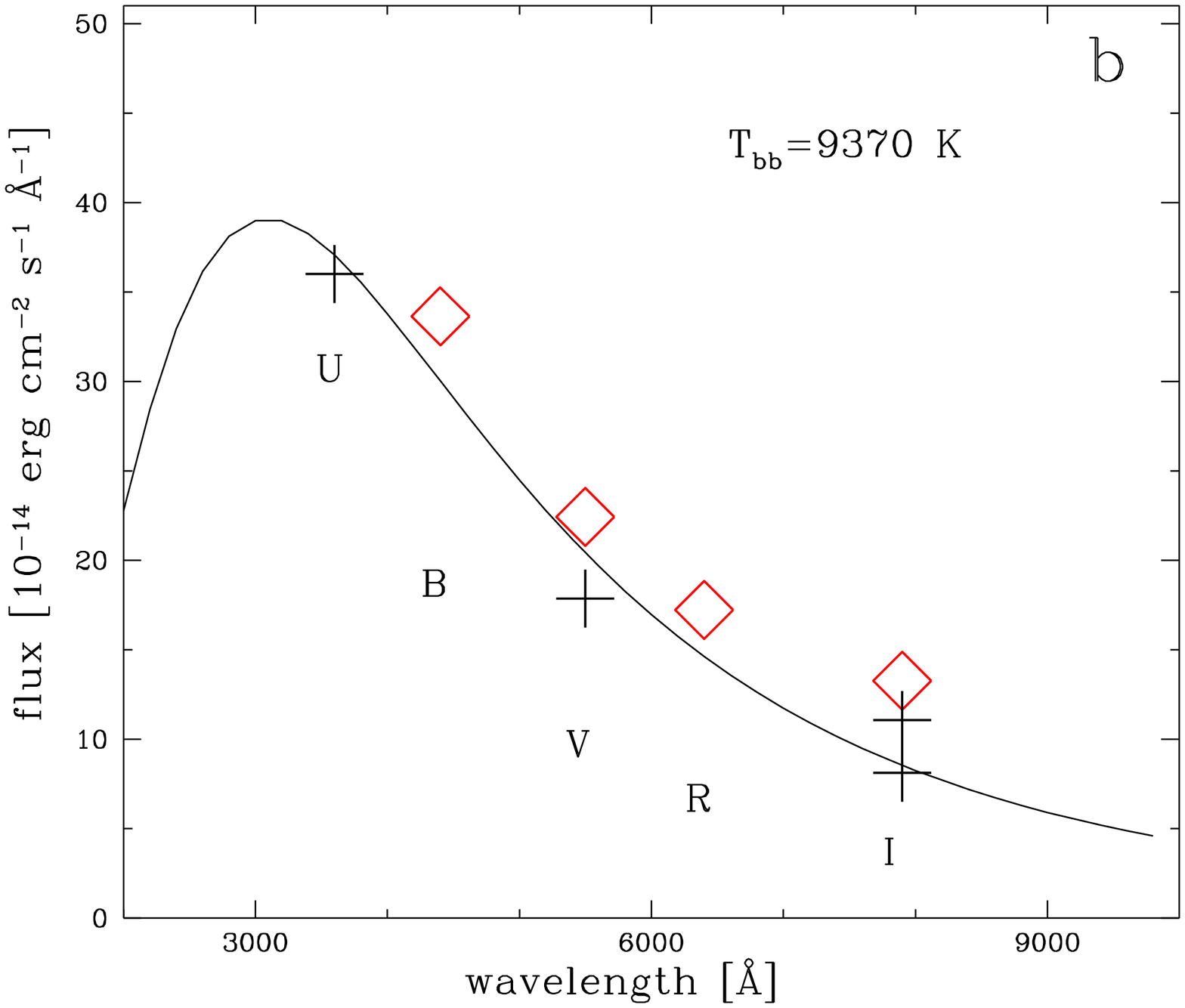}   
  \includegraphics{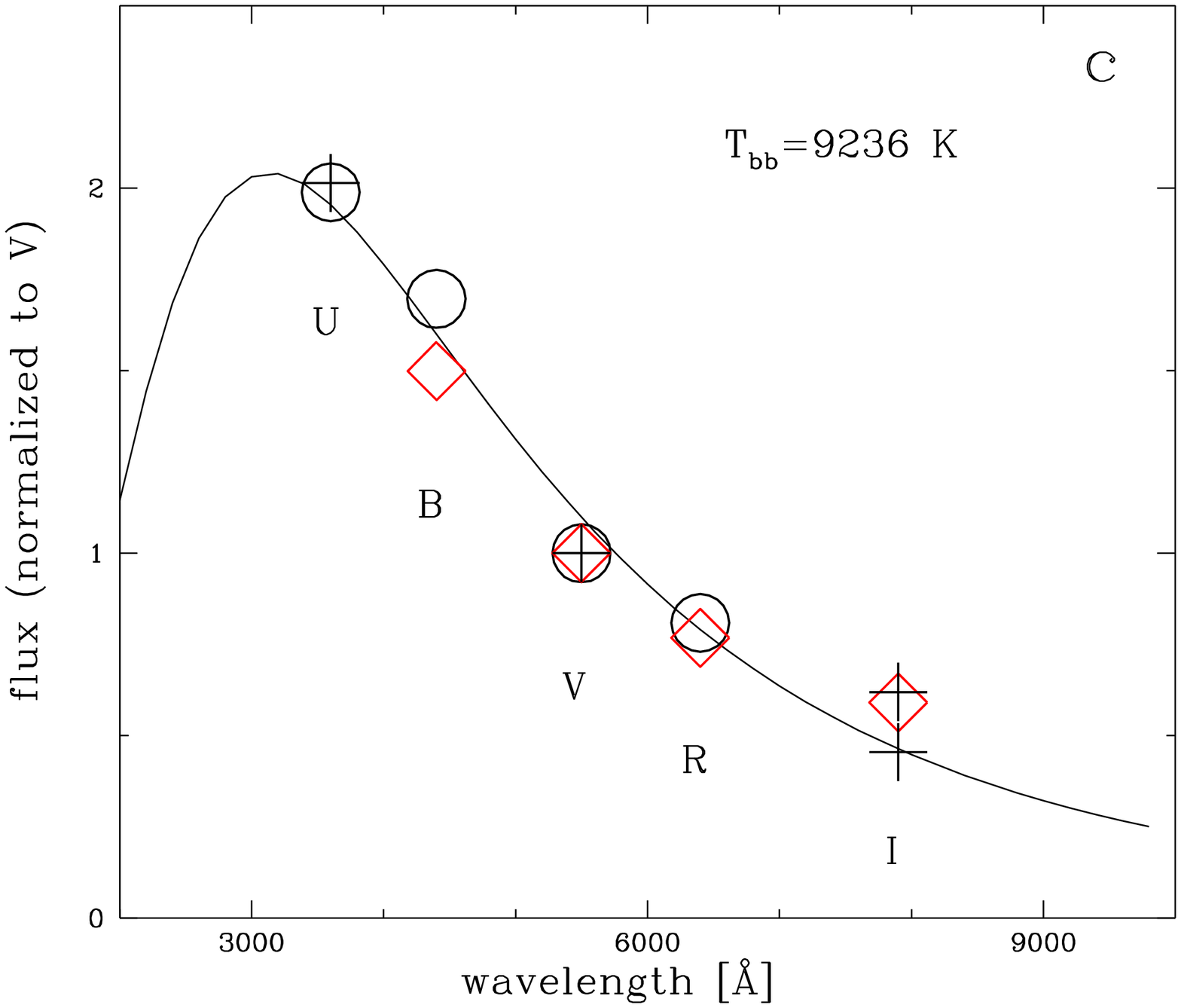}   
  \caption[]{Dereddened fluxes of the flickering light source of RS~Oph. The solid line 
    represents a black body fit. \\
    {\bf a)} 2008 July 6 (circles), $T_{bb}=9492$ K, radius $R=2.6$~\rsun, located at distance 
    $d=1.6$~kpc. \\
    {\bf b)} 2009 July 21 (diamonds) and 2009 July 23 (pluses) plotted together. The fit is
    $T_{bb}=9370$ K, radius $R=4.7$~\rsun.  \\
    {\bf c)} All data normalized to V band  and plotted together. 
     The fit is $T_{bb}=9236$ K. Symbols are identical to those in a) and b).
      }		    
\label{bb}     
\end{figure*}	    

In Fig.\ref{figCOLORS} we plot the colours of RS~Oph versus the $R$ band magnitude for 2008 July 6. 
The $R$ band is chosen because we have the shortest exposure time.
In order to relate magnitudes in different bands, 
taken at non-matching start times and different exposure times, 
we linearly interpolated the light curves. 
The colours have been calculated over 3 grids with time 
resolutions  70, 120 and 360 seconds, 
to match the poorest filter sampling in the corresponding colour relation:
 70~s. for  $(V-R)$ vs. $R$, 
120~s for $(B-R)$ vs. $R$, and 360~s for $(U-R)$ vs. $R$. 
Linear fits (of type $y=a+bx$) to 
the data points in Fig.\ref{figCOLORS} give:
\begin{eqnarray}
(U-R) = -9.73 (\pm1.15) + 1.17(\pm0.11) R   \\
(B-R) = -5.91 (\pm0.71) + 0.79(\pm0.07) R   \\ 
(V-R) = -0.86 (\pm0.29) + 0.17(\pm0.03) R   
\end{eqnarray}
The errors of the coefficients are given in brackets. 
These relations are obtained on the basis of the observations from 
2008 July 6. They  are  valid over the range  $10.20 \le R  \le  10.35$ mag.
The Spearman's (rho) rank correlation gives
$\rho=0.85$  (significance $ 8.10^{-6}$) for Eq.1, 
$\rho=0.66$  ($1.10^{-7}$) for Eq.2, 
$\rho=0.57$  ($2.10^{-10}$) for Eq.3.
The significance in Eq.1-Eq.3 is always $<<0.001$ indicating that all these
correlations are highly significant and changes in the colours of RS~Oph are
correlated with brightness variations.


\subsection{Colours of the flickering source} 
\label{Sec.Col}

Bruch (1992) proposed that the light curve of
CVs can be separated into two parts -- constant light
and variable (flickering) source. We assume that all the variability 
is due to flickering. In these suppositions the flickering 
light source is considered 100\% modulated.  
Following these suggestions,
we calculate the flux of the flickering light source 
as $F_{fl}=F_{av}-F_{min}$, where $F_{av}$ is the average flux 
during the run and $F_{min}$ is the minimum flux during the run
(corrected for the typical error of the observations).
$F_{fl}$ has been calculated for each band, using the values 
given in Table 1 and  Bessel (1979) calibration 
for the fluxes of a zero magnitude star. 

Adopting these results, we find that the flickering light
source contributes about 6\% of the light in the $R$ band, 
7\% in $V$, 10\% in $B$,  and 13\% in $U$.  


Following Snijders (1987) we assume interstellar extinction $E_{B-V} =0.73$
and an extinction law as given in Cardelli et al.(1989).
Using the colour relations (Eq.1--Eq.3, derived in Sect.3.1)
in the interval $R=10.28-10.34$ mag (in other words average $R=10.28$ and we calculate 
average values for other bands, minimal brightness in $R=10.34$, and we derive minimal fluxes 
in other bands), we obtain colours of the flickering light source in RS~Oph as 
$(U-B)_0=-0.67\pm0.06$,
$(B-V)_0=0.08\pm0.06$, 
$(V-R)_0=0.28\pm0.08$.
On 2009 July 21  we obtain  $(B-V)_0=0.22$ and $(V-R)_0=0.22$. 
Using the entire runs in $U$ band from 2009 July 23 and $B$ and $V$  from 2009 July 21 (as plotted in Fig.\ref{bb}) we calculate $(B-V)_0=0.22$ and $(U-B)_0=-0.57$.
The values are similar to the colours of the flickering source in 
1983 July-August $(U-B)_0=-0.71$ and $(B-V)_0=0.07-0.11$
(Bruch 1992).

%
%
%
%
%

\section{Flickering}

\subsection{Temperature and size of the flickering source}
\label{Tfl}

The derived $(U-B)_0$ colour corresponds to a B4V star ($T_{eff}\approx 17000$~K)
and a black body with $T \approx 9000$~K.
The $(B-V)_0$ colour corresponds to an A5V ($T_{eff} \approx 9000$~K)
star and a black body with $T\approx 10000$~K.
$(V-R)_0$  corresponds to a F8V star ($T_{eff}=7000$~K).
These estimates give an approximate temperature of 
the flickering light source $T_{fl}=9000 - 12000$~K.

For the flickering light source we obtain magnitudes 
(corrected for interstellar extinction): 

2008 July 6 : $U = 11.40 $, 
$B = 12.06 $, 
$V = 11.92 $,
$R = 11.55 $ mag;

2009 July 21 :            $B= 10.73$, $V= 10.52 $, $ R= 10.29 $, $ I= 9.94  $ mag;

2009 July 23 : $U=10.00$,             $V= 10.76 $, $ I= 10.11 - 10.45$ mag.

The errors are in $U  \pm 0.04 $,  in $B  \pm 0.05 $, in $V  \pm 0.08 $,
in $R  \pm 0.10 $ in $I \pm 0.12$.

We  adopt $d =1.6\pm0.3$ kpc as given in Bode (1987).
In Fig.\ref{bb} we plot these magnitudes transformed to fluxes. 
Using a black body fit ($nfit1d$ routine of IRAF), we calculate for the flickering light source:
$T_{fl}=9492\pm300$~K; $R_{fl}=2.6\pm0.3$~\rsun; $L_{fl}\sim 50$~$L_\odot$
(2008 July 6)  and  $T_{fl}=9370\pm300$~K; $R_{fl}=4.7\pm0.3$~\rsun;
$L_{fl} \sim 150 \: L_\odot$ (2009 July 21 and 23).
In Fig.\ref{bb}c we plot all points normalized  to  the corresponding V band flux. 
The fit gives $T_{fl}=9236\pm200$~K;

Our $I$-band  data  from 2008 July 6 have considerably larger observational errors 
due to a technical problem which gave as a consequence a variable dark current pattern of the CCD. 
$U$ from 2009 July 21 and $B$ from 2009 July 23 are short. 
They are therefore not used in the black body fits nor plotted in Fig.\ref{bb}. 

There are several different estimates of the mass accretion rate in RS~Oph.
An estimate of  $\dot M_{acc} \approx 2.10^{-7} \: M_\odot$~yr$^{-1}$ arises  
from the required mass accreted on to the WD to give rise 
to the outburst plus the latest inter-outburst time scale of 21 years
(Osborne et al. 2010). 
Assuming $R_{WD} = 0.003$~\rsun\ (Starrfield et al. 1991), this results in
a total accretion luminosity 
$L_{acc}\approx  G M_{wd} \dot M / R_{wd} \approx 2900$~\lsun . 
Alternatively, Walder et al. (2008) explore the hydrodynamics of wind accretion 
in the system and find 
$\dot M_{acc} \approx 10^{-8} \: M_\odot$~yr$^{-1}$, i.e. 
$L_{acc}\approx 146$ \lsun\  
(which compares to $L_{acc}\approx 190 - 740$~\lsun\
from IUE observations at quiescence - Snijders 1987).
The spectral energy distribution of the accretion disk 
around the WD is similar to mid-B spectral type
(Dobrzycka et al. 1996) with luminosity $100-600$~\lsun. 
In practice therefore, the different estimates 
give $L_{acc} \sim 140-2800$~\lsun,
which means that 5\%-35\% of the total accretion luminosity
is  emitted by the flickering source
(in other words $L_{fl}/L_{acc}\sim 0.05 - 0.35$).

\subsection{Time delay}
We searched for time lags between the light curves in different bands, using the
interpolation cross-correlation method (Gaskell \& Sparke, 1986), but found
no delays larger than 10 sec, which is shorter than our time resolution
(the exposure times).

 \begin{figure}
 \mbox{}  
 \vspace{7.2cm}  
  \includegraphics{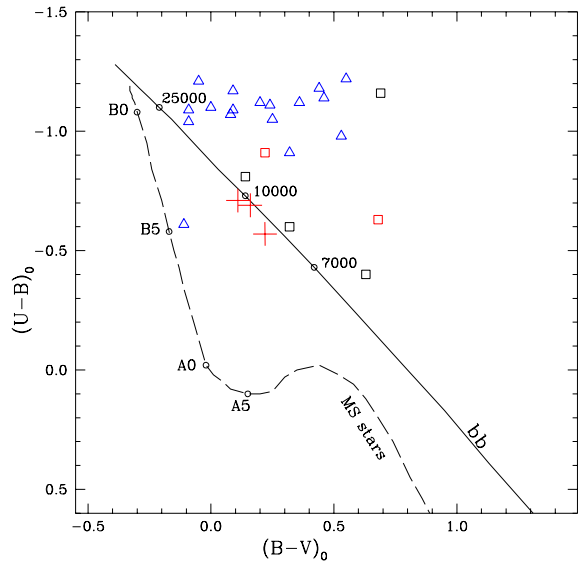}   
  \caption[]{Position of the flickering light 
  source on a $(U-B)$ vs $(B-V)$ diagram.
  The dashed line is the main sequence. The solid line 
  is a black body,  (blue) triangles -- CVs (from Bruch 1992),
  (black) squares -- symbiotic stars (MWC~560, V407 Cyg), RS Oph is marked with (red) plus symbols, 
  T CrB -- (red) squares.
  }		    
\label{2C}     
\end{figure}	    

\subsection{Difference between symbiotics and CVs?}

Fig.~\ref{2C} represent a 2-colour diagram, 
$(U-B)_0$ versus $(B-V)_0$ for the 
flickering light source in a number of CVs and symbiotics. 
The data for the CVs and T~CrB are from Bruch (1992). 
We also plot 2 points for MWC~560 and 2 for V407~Cyg, which are calculated from
the data of  Gromadzki et al. (2006) 

The crosses represent RS~Oph. The three crosses refer to the three epochs of observations -- 1983, 
2008, 2009. They lie well outside the area
of the CVs. 
To address the possible errors of our measurement, we separated 
our light curves of RS~Oph (plotted in Fig.~\ref{fig1-1}) into two parts 
(from UT~19:46 to UT~20:26 and UT~20:27 -- 21:22).
We obtained for the flickering source 
$(U-B)_0=-0.61$  $(B-V)_0=0.11$ for the first time  interval, 
and $(U-B)_0 = -0.74$  $(B-V)_0 = 0.15$  for the second.
This indicates that for RS~Oph the errors are 
approximately equal to the size of the symbols.

To search for differences between the flickering of symbiotic stars
and CVs we performed a two-dimensional Kolmogorov-Smirnov test 
(Peacock  1983; Fasano \& Franceschini 1987),
using the data plotted on Fig.\ref{2C}.
We compare colours of the flickering 
of the recurrent novae RS~Oph and T~CrB with those of the CVs.
The test gives a probability of $2 \times 10^{-3}$ that both distributions are
extracted from the same parent population.

We also performed the same test but comparing CVs with all symbiotic stars.
The test gives a probability of $4 \times 10^{-4}$ that both distributions are
extracted from the same parent population. 
The number of points is not high, but sufficient, 
as shown in Fasano \& Franceschini (1987), and the difference is significant.
This result indicates that in this diagram 
there are clues to the cause of the differences between the flickering 
of these two classes of accreting white dwarfs. 

In the symbiotics of course, the mass donor is a red giant and the 
orbital periods are $>100$ d. In the CVs the mass donors are late-type dwarfs and 
the orbital periods are $\sim 1000$ times shorter.
To the best of our knowledge this is the first evidence that 
flickering differs in these  types of accreting WDs and the physical cause of this difference warrants further investigation (see below).

\section{Discussion}

In quiescence RS~Oph varies irregulary between $V=10.0-12.2$ 
(Collazzi et al. 2009,  AAVSO light curves).
This variability is observed before as well as after the 2006 outburst. 
Darnley et al.(2008) detected an increase of the brightness in $B$ from $B=13.5$ to $B=11.9$ mag 
and from  $V=12.0$ to $V=10.5$  for  a year after the 2006 outburst. 
Worters et al. (2007)  detected an increase in $V$ from $V=11.3$ to $V=11.9$ for 2 weeks. 
The brightness of RS~Oph at the time of our flickering observations is well inside
these limits (unfortunately, there does not seem to be any published $U$ light curve). 
The derived parameters of the flickering can be considered as typical for quiescence. 

After the 2006 outburst the flickering of RS~Oph disappeared (Zamanov et al. 2006)
probably as a result of (1) destruction of the accretion disk 
from the nova explosion  
or (2) a change in the inner disk associated with 
jet production (Sokoloski et al. 2008). 
The flickering resumed by day 241 of the outburst (Worters et al. 2007). 
For the symbiotic star CH~Cyg, Sokoloski \& Kenyon (2003) observed
changes in the flickering in association with the mass ejection event.
Observations of CH~Cyg and RS~Oph therefore indicate 
that a connection does exist between
the flickering behaviour and the ejection of matter from the WD.

Different sites for the origin of the flickering have been discussed.
They are all related to the accretion process: 

{\bf (i)} the bright spot (the region 
of impact of the stream of transferred matter from 
the mass donor star on the accretion disk); 

{\bf (ii)} the boundary layer (between the innermost accretion disk 
and the white dwarf surface); 

{\bf (iii)} inside the accretion disk itself. \\

The findings of Bruch (2000) for HT Cas, V2051 Oph, IP Peg and UX UMa
demonstrated  that the flickering in these CVs can 
originate in both regions {\bf (i)} and {\bf (ii)}.

{\bf (i)} The temperature and the size of the bright
spot are derived for a number of CVs. A few examples of temperatures are: 
for OY~Car Wood et al (1989) calculated 
black body $T=13800\pm1300$~K, and color temperature $T=9000$~K;
 Marsh (1988) for IP~Peg  -  $T = 11200$~K; 
 Zhang \& Robinson (1987) for U~Gem - $T=11600\pm500$~K;
 Robinson et al. (1978) give $T=16000$~K for the bright spot in WZ~Sge.
The temperature of the optical flickering source of RS Oph
$T_{fl} \approx 10000$~K (see Sect.4.1) is similar to the temperature of the 
bright spot for the CVs. 
 The mass transfer in RS Oph may not be from Roche 
lobe overflow (as in CVs), but via stellar wind accretion.  
As a result of the supersonic motion of the WD through
the red giant wind there should be a shock cone and an accretion wake.
Around the WD, it is likely that an accretion disk and/or a cocoon is formed.
In the place where the matter flowing through the accretion wake encounters 
the accretion disk/cocoon a bright spot could be formed (analogous to the bright spot formed in the case of CVs, 
where the mass flow from the inner Lagrangian point encounters 
the accretion disk).

{\bf (ii)} Another possible site for the origin of the flickering in RS~Oph 
is the boundary layer between the white dwarf and accretion disk.
Bruch \& Duschl (1993) estimated the size of the boundary layer ($\epsilon$) in RS~Oph as
$\epsilon = 2.20$. In their model 
$L_{d}/L_{bl}$  is connected with the size of the boundary layer
($L_{d}$ is the the luminosity of the accretion disk,  $L_{bl}$ 
is the  luminosity of the boundary layer; see Fig~1 in Bruch \& Duschl 1993). 
If we assume $L_d + L_{bl} \approx L_{acc} \approx 110-2000$ \lsun, and 
$L_{bl} \approx L_{fl} \approx 5 - 150$ \lsun, then
we calculate $L_{d}/L_{bl}=1.2-12$, and the lower value
agrees with the size of the boundary layer as estimated by Bruch \& Duschl (1993).
However, $T_{fl}$ of RS~Oph as derived in Sect. 4.1
is too low for a boundary layer.

{\bf (iii)} For UU~Aqr, Baptista \&  Bortoletto (2008) found no evidence of 
flickering generated  in regions {\bf (i)} and {\bf (ii)}.
 They suggested that the 
flickering in UU Aqr is generated in the accretion disk itself and a possible 
reason can be turbulence 
generated after the collision of disk gas with the density-enhanced
spiral wave in the accretion disk.

The radial temperature profile of a steady-state accretion disk is:
\begin{equation}
T_{eff}^4=\frac{3 G \dot M_{acc} M_{wd}}{8 \pi \sigma R^3} 
\left[ 1-\left(\frac{R_{wd}}{R}\right)^{1/2} \right] ,
\end{equation}
where $\sigma$ is the Stefan-Boltzmann constant, $R$ is the radial distance from the WD.
Using the parameters for RS~Oph, 
a temperature  $T_{fl} \sim$9500 K 
(the temperature of the flickering light source as derived in  
Sect. \ref{Tfl}) should be achieved at a distance 
$R\approx 0.5-1$ \rsun\ from the WD.
If {\bf (iii)} is the place for the origin of the flickering of RS~Oph, then it 
comes from $R$~\simlt~1~\rsun\ from the WD.

To understand more fully the nature of the flickering  variability  of RS~Oph
we need to acquire a set of spectra simultaneously with 
photometry and with time resolution $\sim 30$ seconds (see also Sokoloski 2003). 
Such spectra potentially can directly give the spectrum of the 
flickering light source. 

\section{Conclusions}

We report our CCD observations of the flickering variability of 
the recurrent nova RS Oph, simultaneously with 4 telescopes in the
$UBVRI$ bands.  RS~Oph has a flickering source with 
$(U-B)_0=-0.62 \pm 0.07$, 
$(B-V)_0=0.15 \pm 0.10$, 
$(V-R)_0=0.25 \pm 0.05$.

For the flickering light source in RS~Oph 
(1) we estimate 
$T_{eff} \approx 9500 \pm 500 $~K, which is similar to the temperature 
of the bright spot in CVs;
(2) using a distance of $d=1.6$~kpc, we find 
size $R_{fl}\approx 3.5\pm0.5$~\rsun, and luminosity $L_{fl}\sim 50 - 150~L_\odot$.
 
We find that on the $(U-B)$ vs $(B-V)$ diagram the flickering of 
the symbiotic stars  differs from the flickering of the 
cataclysmic variables. 
  
The possible sites for the origin of the  flickering are briefly discussed and it is suggested that time-resolved spectroscopy should be carried out to explore this in more detail.
  
\section*{Acknowledgments}  
RKZ, AG, and KAS acknowledge the partial support by Bulgarian NSF (HTC01-152)
and Slovenian Research Agency (BI-BG/09-10-006).
We are grateful to the referee, Dr. P.~Woudt, for very helpful comments on the initial version of this paper.

\end{document}